\documentclass[aps,prl,twocolumn,superscriptaddress,showpacs,english]{revtex4-1}

\usepackage[T1]{fontenc}
\usepackage[latin9]{inputenc}
\usepackage{babel}
\usepackage{amsmath}
\usepackage{amssymb}
\usepackage{wasysym}
\usepackage{graphicx}
\usepackage{xcolor}
\usepackage{bm}          
\usepackage{xspace}

\definecolor{mydarkgreen}{rgb}{0.0,0.5,0.0}
\definecolor{friebrick}{rgb}{0.698,0.1333,0.1333}
\usepackage[linktocpage=true,
  colorlinks=true, 
  pdfborder={0 0 0},
  linkcolor=blue,
  citecolor=friebrick,
  filecolor=yellow,
  urlcolor=mydarkgreen,
  bookmarks,
  pdfauthor={},
]{hyperref} 

\newcommand{\sh}{H$_3$S}

\newcommand{\tc}{T$_{\text{c}}$}

\newcommand{\QE}{{\textsc{Quantum ESPRESSO}}\xspace}

\newcommand{\cubic}{$Im\overline{3}m$}

\newcommand{\Riken}{RIKEN Center for Emergent Matter Science, 2-1 Hirosawa, Wako, 351-0198, Japan}
\newcommand{\UniRoma}{Dipartimento di Fisica, Universit\`a di Roma La Sapienza, Piazzale Aldo Moro 5, I-00185 Roma, Italy}
\newcommand{\UniTokyoPhy}{Department of Applied Physics, University of Tokyo, Tokyo 113-8656, Japan}
\newcommand{\USendai}{Department of Physics, Tohoku University, 6-3 Aza-Aoba, Sendai, 980-8578 Japan}

\begin{document}


\title{Absence of conventional room temperature superconductivity at high pressure in carbon doped H$_3$S}

\author{Tianchun Wang}
\email{tianchun.wang@riken.jp}
\affiliation{\UniTokyoPhy}
\author{Motoaki Hirayama}       
\affiliation{\UniTokyoPhy} 
\affiliation{\Riken}
\author{Takuya Nomoto}          
\affiliation{\UniTokyoPhy}
\author{Takashi Koretsune}      
\affiliation{\USendai}
\author{Ryotaro Arita}          
\affiliation{\UniTokyoPhy} 
\affiliation{\Riken} 
\author{Jos\'e A. Flores-Livas} 
\email{jose.flores@uniroma1.it}
\affiliation{\UniRoma}     
\affiliation{\Riken} 
\date{\today}

\begin{abstract}
In this work, we show that the same theoretical tools that 
successfully explain other hydrides systems under pressure seem to be at odds with the recently claimed
conventional room temperature superconductivity of the carbonaceous sulfur hydride. We support our conclusions with I) the absence of a dominant low-enthalpy stoichiometry and crystal structure in the ternary phase diagram. II) Only the thermodynamics of C-doping phases appears to be marginally competing in enthalpy against H$_3$S.  
III) Accurate results of the transition temperature given by {\it ab initio} Migdal-Eliashberg calculations differ by more than 110\,K to recently theoretical claims explaining the high-temperature superconductivity in carbonaceous-hydrogen sulfide. A novel mechanism of superconductivity or a breakdown of current theories in this system is possibly behind the disagreement. 
\end{abstract}

\maketitle


Over the last decade, pressurized hydride compounds have led the path to many important landmarks in superconductivity. Notable cases include silane in 2008~\cite{Eremets_silane_2008}, H$_3$S in 2014~\cite{DrozdovEremets_Nature2015,Einaga_H3S-crystal_NatPhys-2016} which has triggered most of the field, and the confirmation of high-\tc\ in LaH$_{10}$ by independent teams in 2019~\cite{Hemley-LaH10_PRL_2019,Nature_LaH_Eremets_2019,Errea_Nature_2020}. This unfolding success of important breakthroughs is largely due to the symbiosis of theory, computation, and experimental sciences, which has accelerated the discovery by pointing to niches of interesting systems~\cite{Flores_review,BI2019,pickard2020superconducting,zhang2017materials,oganov2019structure}. 

Recently, Snider {et al.}~\cite{Snider2020} achieved a decades-old quest; they reported solid evidence of the first room-temperature superconductor (RTS) made of carbon-sulfur-hydrogen. Although the report set a landmark in the {\it annals} of science, there are still many open questions surrounding this important discovery. 
For instance, the exact stoichiometry of the claimed carbonaceous hydrogen sulfide that exhibits RTS is still elusive. 
Moreover, there is a debate with confronted arguments on the possibility of unusual superconducting features in all superhydrides at odds with the Bardeen-Cooper-Schrieffer theory~\cite{hirsch2020absence,hirsch2021absence,Hirsch_PRB_2021}. It includes the sharp drop of electric resistivity at \tc\ and its dependence on a magnetic field~\cite{hirsch2020absence,talantsev2021electron,dogan2021anomalous}.
It is worth noticing that the room-temperature superconductor reported at 287.7\,K at 267\,GPa, has not been confirmed by magnetic susceptibility measurements~\cite{Snider2020}. 
But, amidst such unsolved puzzles, perhaps the most intriguing question is: what is the crystalline structure of the RTS? 


Certainly, it is difficult to measure the crystalline structure of a tiny sample under extremely high pressure, additionally complicated by the small scattering ratio of low-Z hydrogen. Hence, to clarify the mechanism of superconductivity and electronic and phonon properties, it is highly desirable to know the crystalline structure from the theoretical side. So far, there have been two works on crystal structure prediction for C-S-H ternary systems~\cite{Sun2020,Cui2020}. In these works, plausible structures were explored under high-pressure $p=100$ GPa~\cite{Sun2020} and 100-200\,GPa~\cite{Cui2020}, which reported many structure candidates for high-\tc\ superconductivity, including CSH$_7$. However, these candidates are {\it not} thermodynamically stable, and also, the pressure explored is much lower than $p\sim$270\,GPa, at which the RTS was reported.

In this Letter, we shed light on different open issues of the RTS. 
Resorting to structure prediction, we enlarged the chemical composition search and estimated C-S-H ternary systems' formation enthalpy at 250\,GPa. We also analyzed the doped phases, from their thermodynamic stability to superconducting properties and found flagrant differences compared to recent theoretical reports on C-doped \sh~\cite{Ge_2020,hu2020carbondoped}. On the transition temperatures obtained within virtual crystal approximation (VCA) and McMillan-Allen-Dynes (MAD) theory, presumably, an electronic smearing parameter holds to overestimate \tc 's theoretical value. In contrast, we found more than 110\, K difference in \tc\ between experimental and the converged {\it ab initio} values. Our theoretical results show that the room-temperature superconductor cannot be explained by conventional superconductivity of carbon-doped H$_{3}$S phases nor other stoichiometries explored so far. 



\begin{figure*}[t!]
\includegraphics[width=1.9\columnwidth,angle=0]{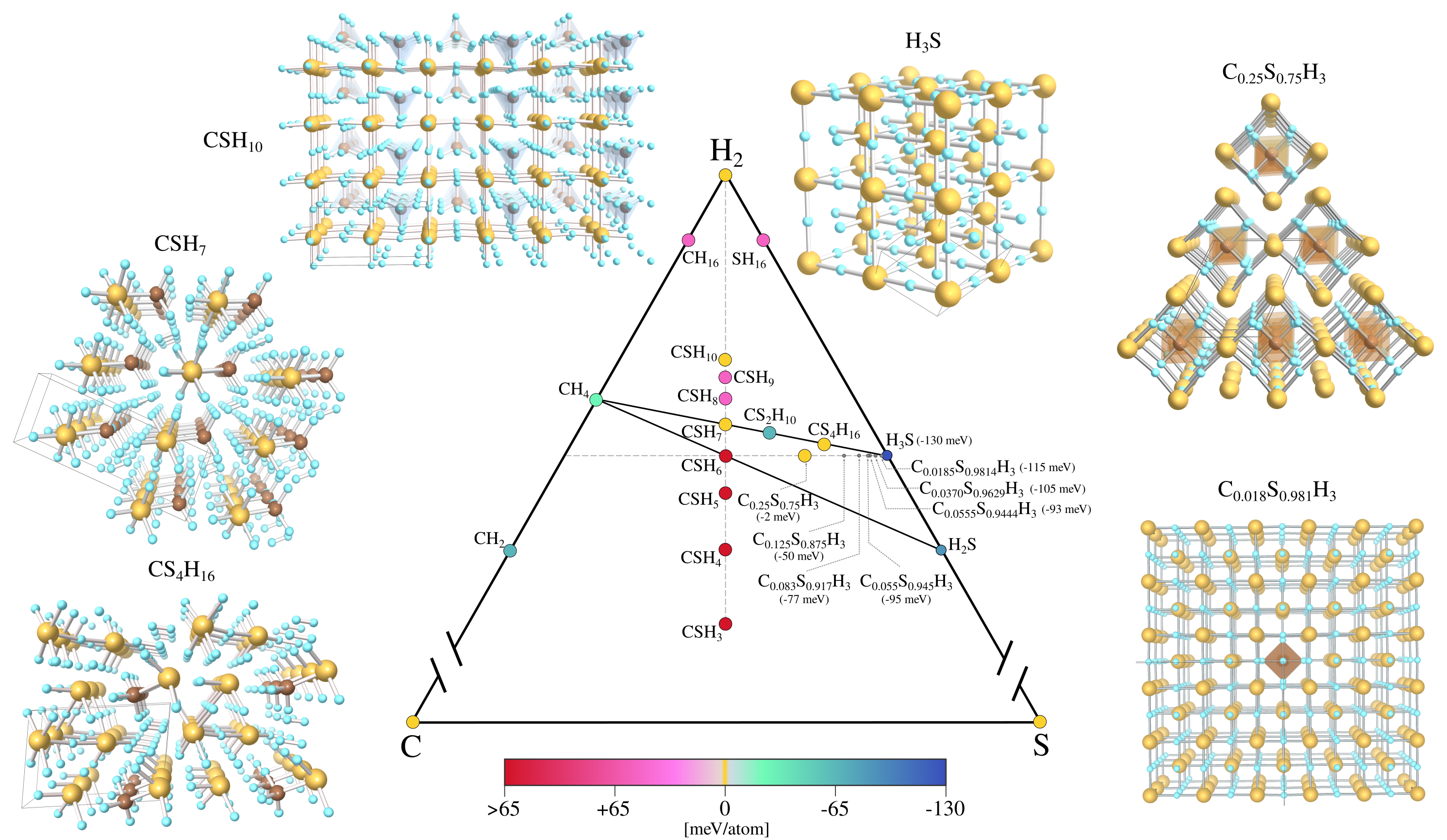}
\caption{~C-S-H convex hull of the formation enthalpy calculated for selected stoichiometries at 250\,GPa. Elemental phases and other low-lying enthalpy compositions are coloured in yellow, in red, unstable, and from blue to dark-blue, chemical compositions with negative formation enthalpy. At this pressure, the lowest enthalpy phase is H$_{3}$S (130\,meV), followed by doped phases. However, already at 3.7~\% of C incorporation, the enthalpy difference between the parental phase and the dope one changes by 25\,meV or $\sim$290 K. Representative low-enthalpy structures of selected compositions are displayed outside the ternary hull. Intriguingly, in all the different regions studied in the phase diagram, the motifs with the lowest enthalpy correspond to molecular parts, either H$_x$C and H$_x$S, but nothing points towards a {\it fused}, covalently bonded C-S-H compound.}
 \label{fig:Hull}
\end{figure*}

Due to the prohibited computational overhead of calculating a huge number of available compositions in the ternary system, we focus our strategy on exploring only representative sections of the compositional landscape in detail. Fig.~\ref{fig:Hull} shows the C-S-H ternary convex hull for selected compositions at 250\,GPa (see details in Ref.~\footnote{The compositional and configurational space of C-S-H was explored using the minima-hopping method~\cite{Goedecker_mhm_2004,Amsler_mhm_2010,flores-sanna_PH3_2016,flores2020crystal}. The representative compositions of the ternary phase diagram were carried out with different formula units (containing up to 48 atoms) and carried out under pressure. Energy, atomic forces, and stresses were evaluated using the density-functional theory (further details are included in the Supplemental Information).}).  
We find CSH$_{7}$ (enthalpy of formation $\sim$0 meV) and the absence of a dominant (low enthalpy) phase, which is in agreement with previous works~\cite{Sun2020,Cui2020}. In our searches, we observed that different sections of the compositional space were governed by anticipated trends. 

The high content of carbon and hydrogen (top left areas of the Gibbs triangle) will form CH$_{2}$ and CH$_{4}$. Increasing hydrogen content (H$_{5-16}$) in these areas will then produce phase separation to H$_{2}$ and CH$_{x}$, which are compositions with formation enthalpy well above 100\,meV/atom (not shown). Moving to the middle section of the triangle, for C and S on 1:1 ratio with increasing H; H$_{3}$, H$_{4}$, H$_{5}$ and H$_{6}$; these stoichiometries are highly energetic and unlikely to occur. In these phases, decompositions to H$_{2}$, H$_{3}$S or CH$_{x}$ are seen. 

\begin{figure*}[t!]
\includegraphics[width=1.9\columnwidth,angle=0]{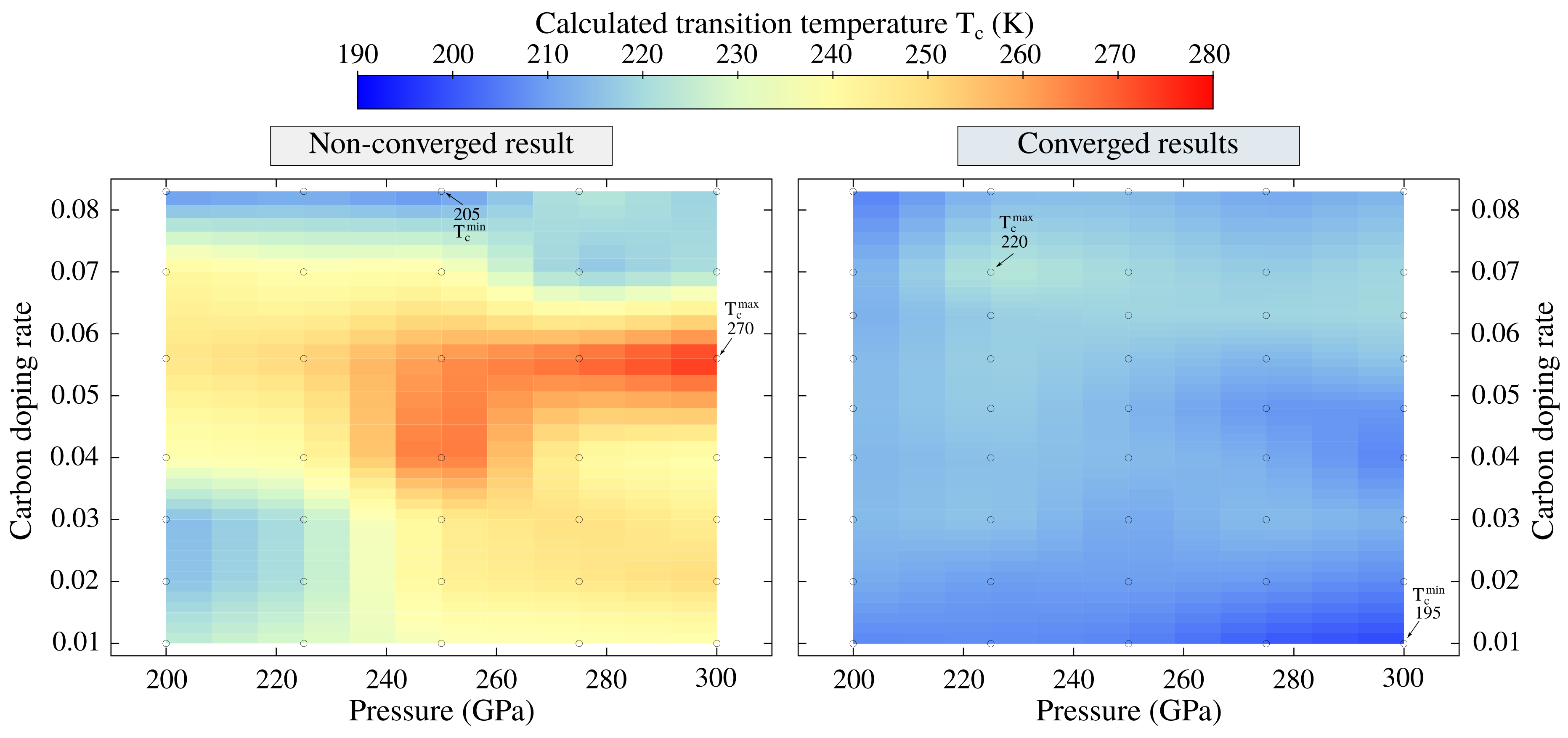}
\caption{~Calculated doping-superconductivity-pressure phase diagrams using two protocols: the left panel shows \tc\ with a non-converged parameter (see text) and the right panel with the controlled and converged protocol. The abscissa in both panels displays pressure in the range where the RTS was reported. The ordinate represents the carbon doping as simulated by the VCA and the colour bar shows the estimated \tc\ given by the MAD-formula. The Coulomb parameter $\mu^*$ is set to 0.1. Maximum and minimum \tc 's for each panel are shown. Beyond the technical validity of the VCA, it is clear that independently of doping and pressure, an RTS is absent for the converged case. }
 \label{fig:VCA}
\end{figure*}

CSH$_{7}$ and CSH$_{10}$ are interesting compositions that become metastable due to their conformation and "poor" metallicity. These compositions are formed by H$_{3}$S and CH$_{x}$ units with weak Van der Walls interaction~\cite{Sun2020} between them (detrimental to high-\tc\ phonon-superconductivity). Increasing the hydrogen above H$_{10}$ in the ternary compounds, C-H binaries or S-H binaries also results in phase separation. Most of the found phases present simple patterns, and these can be classified almost as amorphous phases. Coming back to the ratio of C$_{0.25}$S$_{0.75}$, the same pattern emerges, shown in the plotted figure (C$_{0.25}$S$_{0.75}$H$_{3}$; the characteristic cubic arrangement of sulfur-hydrogen with distinctive layers and enclosed CH$_{4}$ units (a poor metal).  Moving to C$_{0.5}$S and lower hydrogen (below H$_2$) content seems odd for high-\tc\ superconductivity since decreasing the hydrogen content reduces the chances of finding key ingredients: metallic phases with light atomic mass. Explorations below the CSH$_{3}$ range with C points towards the C-C formation of stable covalent bonds; however, these are semiconducting phases. In sulfur-rich areas, S-S metallic phases are found. However, these are unlikely to be responsible for the RTS. 

The region close to the lowest enthalpy (H$_{3}$S) phase is the most relevant in the compositional space. 
The varying C doping into the matrix of H$_{3}$S (C-substitution in S sites) is the most reasonable solution from the thermodynamical point of view: we report that enthalpy decreases from -2 meV/atom at 25\%C to -50 for 12.5\%C, to -77 meV/atom for 8.3\%C, to -95 meV/atom for 5.5\%C, and finally to -105\,meV/atom for 3.7\%C, and so on until reaching the H$_{3}$S with enthalpy of -130 meV/atom (clearly the dominant composition).
Judging by the convex hull of stability, C-doping offers a possible structure model to explain the RTS, which is not new for other systems (see Ref.~\cite{Flores_review} for a review). Nevertheless, introducing carbon into the H$_{3}$S lattice comes at a price: it plays a detrimental role to single-phase stability (at 3.7\,\% of C doping level, the enthalpy difference between the parental phase and the doped one changes by 25\,meV or $\sim$290\,K), and excessive doping could also worsen the pristine electronic structure of H$_{3}$S. 

It has been reported that, at least in the other two major systems (H$_3$S and LaH$_{10}$), the highly symmetric arrangements of atoms in hydrides under pressure display a van Hove singularity (VHS) near the Fermi level ($E_{\rm F}$)~\cite{Flores_review,LaAlH10_Rome-Tokyo}. 
In the case of H$_{3}$S (close to the C-S-H case), the VHS peak resides slightly lower than $E_{\rm F}$~\cite{PRB_VHA_Akashi}. From the electronic point of view, it is favourable for superconductivity to attempt electron doping. Recent studies based on the McMillan-Allen-Dynes (MAD)~\cite{McMillanTC_PR_1968,AllenDynes_PRB1975} formula have shown exceptionally that \tc\ of C$_x$S$_{1-x}$H$_3$ can be as high as room-temperature when $x$ is $\sim$0.05~\cite{Ge_2020,hu2020carbondoped}. However, it is also well documented that the MAD formula is not a good approximation when the electron-phonon coupling is strong~\cite{AllenMitrovic1983} 
or when the density of states (DOS) has a significant energy dependence around the VHS (as for electron doping)~\cite{AllenMitrovic1983}.  

Let us first examine the effects of carrier doping onto the possible explanation of the RTS. 
Fig.~\ref{fig:VCA} confronts two phase diagrams of doping-pressure-\tc : the left one reproduces satisfactorily the one presented by Ge {et al.}~\cite{Ge_2020}. 
The right panel (this work) shows quite the opposite phase diagram, with much lower \tc\ values. Noticeable, when calculating the Eliashberg function $\alpha^2F(\omega)$, a sensitive parameter is the broadening width $\delta$ of the smearing for the double-delta integral. The left diagram in Fig.~\ref{fig:VCA} is the result using the broadening width $\delta = 0.002$ Ry, and the right one is the result produced by $\delta =$ 0.014 Ry. Since the results have significant dependence on the broadening width $\delta$, to reach convergence, we choose the value of $\delta$ so that it can reproduce $N(0)$, where $N(0)$ is the DOS at the Fermi level given by the tetrahedron method~\cite{Blochl_1994} using a sufficiently dense mesh~\cite{Koretsune_2017}. We ascribe the difference between the plots from a lack of convergence in the article by Ge {et al.}~\cite{Ge_2020}; \tc\ is overestimated, especially with the MAD formula when too narrow smearing is used for the integral of the electron-phonon line width in the momentum space. The right panel summarizes \tc\ using a protocol and carefully tested electronic parameters~\cite{Wang2020} that reproduce theoretical values for H$_3$S~\cite{Sano2016} and LaH$_{10}$~\cite{Errea_Nature_2020}.

\begin{figure}[t!]
\includegraphics[width=0.99\columnwidth,angle=0]{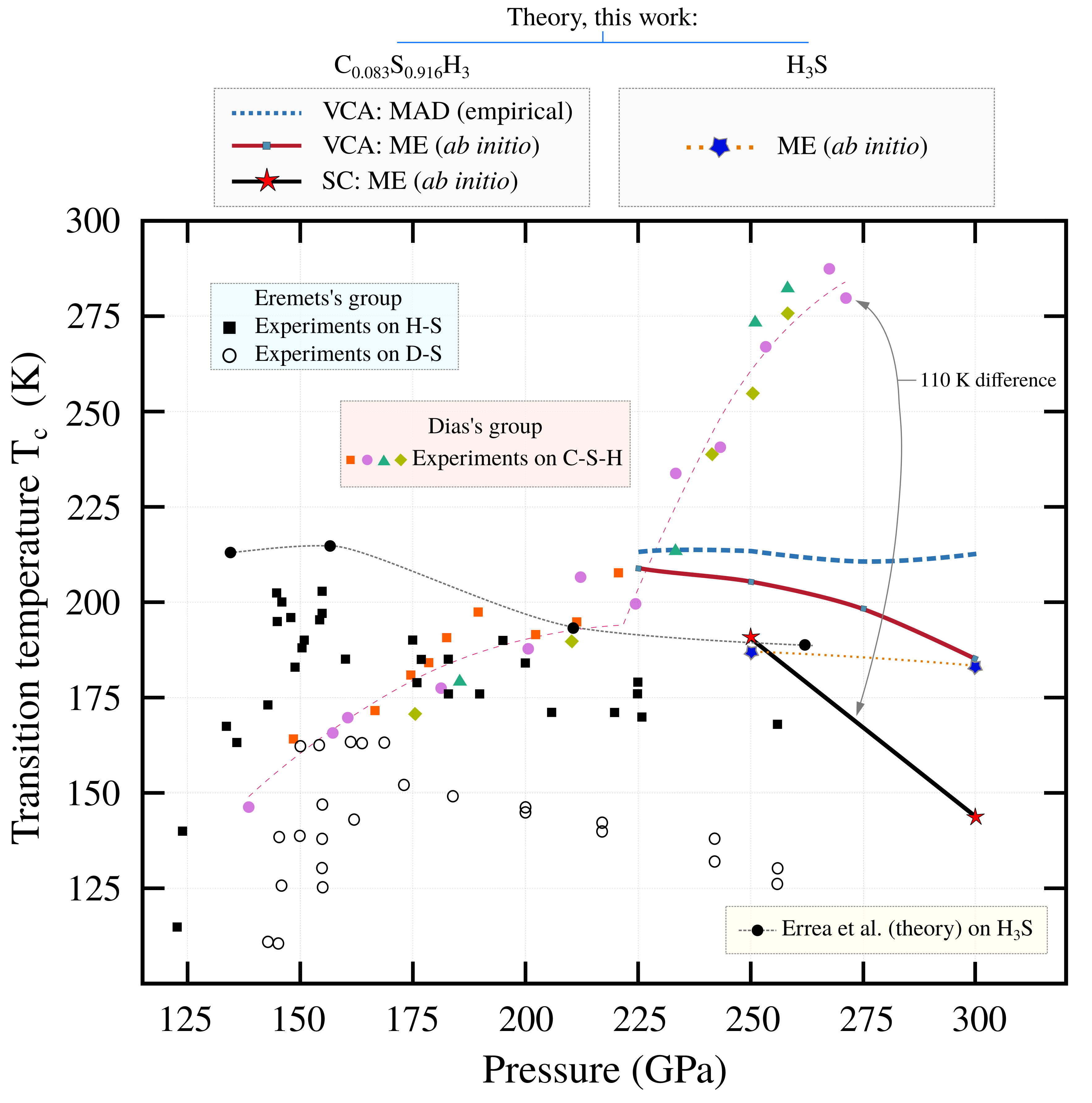}
\caption{~\tc\ vs pressure: Theoretical results (this work) are estimated using different methods. Experimental results reported by Dias's group on C-S-H~\cite{Snider2020}, Eremets's group on H-S and D-H~\cite{DrozdovEremets_Nature2015,minkov2020boosted,mozaffari2019superconducting} and theoretic results on H$_3$S~\cite{Nature_Errea_2016} are also shown. Independent of the methodology used, our results suggest a sizable deviation as large as 110\,K between the most reliable theoretical estimations and experiments on \tc\ (see details in Ref.~\cite{Note2} and Supplemental Information).}
 \label{fig:Tc}
\end{figure}

Besides the discussions above, another shortcoming of the MAD-formula is the Coulomb interaction, which is introduced phenomenologically by a pseudo Coulomb parameter, $\mu^*$, with a value set around 0.1. However, there is no reason why these values should be transferred at high pressure.  In our case, the Migdal-Eliashberg (ME) calculation used the Coulomb interaction kernel, in which we solved the gap equation directly to get rid of $\mu^*$. Thus, the ME calculation is more robust and straightforward, without any empirical parameters involved~\cite{Wang2020}.
Using accurate first-principles Migdal-Eliashberg calculations~\cite{Sano2016,Wang2020} 
(see details in \footnote{To study the doping effect of C atoms in C$_x$S$_{1-x}$H$_3$, 
we used I) the virtual crystal approximation (VCA)~\cite{Nordheim_1931,Bellaiche&Vanderbilt_2000} and performed a thorough examination of C$_x$S$_{1-x}$H$_3$ with $x$ ranging from $1\%$ to $8.3\%$, in which for the virtual atom we use the potential $V_{\rm VCA} = x V_{\rm C}+(1-x)V_{\rm S}$; II) the actual supercell calculation for C atoms doping with 8\% ratio. We use \QE~\cite{Quantumespresso_2017} for the DFT and DFPT calculations~\cite{Baroni_DFPT_RMP2001}, with the exchange-correlation functional proposed by Perdew, Burke, and Ernzerhof~\cite{PBE_PRL1996}. For C and S atoms, the optimized norm-conserving Vanderbilt pseudopotential (ONCVPSP)~\cite{Hamann_2013} was employed. For H atom, we use the ultrasoft pseudopotentials~\cite{Vanderbilt_1990} provided in PSLibrary~\cite{PSLibrary_1.0.0}. To generate $V_{\rm VCA}$, we use the tool \textit{virtualv2.x} implemented in \QE. For the DFT calculation, the cutoff energy of the plane waves for wavefunction expansion is set to be 80 Ry, and cutoff of the charge density is 320 Ry. For calculations using $V_{\rm VCA}$, we use an 18$\times$18$\times$18 $\bm{k}$-mesh and a 9$\times$9$\times$9 $\bm{q}$-mesh for the phonon calculation and a 36$\times$36$\times$36 $\bm{k}$-mesh for the electron-phonon calculation. For the electron-phonon calculation and the Eliashberg calculation, the electronic eigenenergies and wave-functions are calculated using a 36$\times$36$\times$36 $\bm{k}$-mesh. We have also performed supercell calculation for C$_x$S$_{1-x}$H$_3$ with $x=0.083$, using a 3$\times$2$\times$2 supercell. The supercell calculation is performed on a 6$\times$6$\times$6 $\bm{k}$-mesh and a 3$\times$3$\times$3 $\bm{q}$-mesh for the DFPT calculation, and a 12$\times$12$\times$12 $\bm{k}$-mesh is employed for the electron-phonon interaction and Eliashberg calculation. In the Eliashberg calculation, we use an RPA-type static Coulomb kernel, with a 6$\times$6$\times$6 $\bm{k}$-mesh and a 3$\times$3$\times$3 $\bm{q}$-mesh for the Coulomb calculation.}), for which the doping effects are described by the VCA, we found that \tc\ of the doped phases of H$_3$S are hardly {\it enhanced} by $\sim$20\,K and decreases as a function of pressure (red line shown in Fig.~\ref{fig:Tc}). 
This tendency is also observed in H$_3$S~\cite{flores-sanna_HSe_2016,Akashi2015,Akashi2016,Nature_Errea_2016} and  LaH$_{10}$~\cite{Errea_Nature_2020}.

In addition to the shortcomings of the MAD formula used in Fig.~\ref{fig:VCA}, another concern is the validity of VCA when mixing atomic potentials for non-neighbouring species. The theoretical accuracy of VCA to estimate doping phase diagrams of hydrides under pressure is beyond the scope of this work. 
However, we would like to briefly discuss and compare the VCA and the supercell calculation, a more conventional scheme to treat the doping simulations.
The VCA calculation is performed in the primitive unit cell (4 atoms) and depends enormously on the atomic potential, alchemically constructed via a single virtual atom. For systems with a low doping ratio, the supercell contains $50$~to~$100$ atoms. Compared to VCA, the supercell calculation is computationally costly, especially for the electron-phonon estimation. Nevertheless, it is presently the most accurate method for treating doping since it can capture electronic features and crystal symmetry breaking. 

Fig.~\ref{fig:Tc} summarizes the up-to-date available values of \tc\ from experiments (Snider {et al.}~\cite{Snider2020}) and theory (this work).
We compare the supercell calculation with the one using the VCA approximation for estimations with C atom substitution. 
In Fig.~\ref{fig:Tc}, lines indicate the \tc\ dependence upon pressure calculated in this work with different levels of theory, from VCA (MAD-formula), VCA (ME) and supercell (ME). It is noticeable that the level of theory employed does not play a role in describing what one could consider as an RTS. While the results given by the three methods are qualitatively consistent, it confirms the absence of RTS in carbon-doped phases of H$_3$S. In the same figure, we include the ME-\tc\ for H$_3$S and C$_x$S$_{1-x}$H$_3$ with $x=8.3\%$. At 250\,GPa, \tc\ is marginally increased by only $\sim 5$ K, and at 300\,GPa, \tc\ for the C-doped phase is even lower than that of H$_3$S. The enhancement of superconductivity by C atoms, which has lighter mass, is not significant because C$_x$S$_{1-x}$H$_3$ with a low doping ratio has a similar electronic structure and electron-phonon coupling characteristics as compared to H$_3$S. We refer the readers to the Supplemental Information for extensive details on the electronic band structures, phonon density of states, and a thorough investigation of different doping models and how the electronic singularity is altered by carbon.  

Finally, we reach two contradicting points: on the one hand, doping appears to be the most likely explanation for the RTS. However, a low degree of doping does not alter the electronic structure, and \tc\ is close to the reported \tc\ of H$_{3}$S, as shown in Fig.~\ref{fig:VCA}. And on the other hand, large doping alters the so-called {\it fine-tuning}~\cite{Snider2020} of the VHS drastically. 

In summary, from the thermodynamic perspective, substituting carbon in sulfur sites or interstitial space increases the formation enthalpy (it becomes less stable). 
Introducing carbon in the \cubic\ phase of H$_3$S plays against high-\tc , changes the shape of DOS, decouples phonons, slightly modify the lattice and factors down \tc .  
Perhaps the current level of theory is insufficient to reconcile the scenario with the present experimental results. 
We conclude by asserting that in previous systems (H$_3$S~\cite{Nature_Errea_2016} and LaH$_{10}$~\cite{Errea_Nature_2020}), remarkable compatibility between theoretical and experimental sides in \tc\ and phase diagram is found; for the carbonaceous-sulfur hydride, this might not be the case.  

\begin{acknowledgments} 
We would like to thank M. Eremets for fruitful discussions. 
This work was supported by a Grant-in-Aid for Scientific Research (No. 19K14654 and  No. 19H05825), "Program for Promoting Researches on the Supercomputer Fugaku" from MEXT, Japan.
\end{acknowledgments}

\bibliographystyle{apsrev4-1}
\bibliography{main}

\begin{thebibliography}{50}%
\makeatletter
\providecommand \@ifxundefined [1]{%
 \@ifx{#1\undefined}
}%
\providecommand \@ifnum [1]{%
 \ifnum #1\expandafter \@firstoftwo
 \else \expandafter \@secondoftwo
 \fi
}%
\providecommand \@ifx [1]{%
 \ifx #1\expandafter \@firstoftwo
 \else \expandafter \@secondoftwo
 \fi
}%
\providecommand \natexlab [1]{#1}%
\providecommand \enquote  [1]{``#1''}%
\providecommand \bibnamefont  [1]{#1}%
\providecommand \bibfnamefont [1]{#1}%
\providecommand \citenamefont [1]{#1}%
\providecommand \href@noop [0]{\@secondoftwo}%
\providecommand \href [0]{\begingroup \@sanitize@url \@href}%
\providecommand \@href[1]{\@@startlink{#1}\@@href}%
\providecommand \@@href[1]{\endgroup#1\@@endlink}%
\providecommand \@sanitize@url [0]{\catcode `\\12\catcode `\$12\catcode
  `\&12\catcode `\#12\catcode `\^12\catcode `\_12\catcode `\%12\relax}%
\providecommand \@@startlink[1]{}%
\providecommand \@@endlink[0]{}%
\providecommand \url  [0]{\begingroup\@sanitize@url \@url }%
\providecommand \@url [1]{\endgroup\@href {#1}{\urlprefix }}%
\providecommand \urlprefix  [0]{URL }%
\providecommand \Eprint [0]{\href }%
\providecommand \doibase [0]{http://dx.doi.org/}%
\providecommand \selectlanguage [0]{\@gobble}%
\providecommand \bibinfo  [0]{\@secondoftwo}%
\providecommand \bibfield  [0]{\@secondoftwo}%
\providecommand \translation [1]{[#1]}%
\providecommand \BibitemOpen [0]{}%
\providecommand \bibitemStop [0]{}%
\providecommand \bibitemNoStop [0]{.\EOS\space}%
\providecommand \EOS [0]{\spacefactor3000\relax}%
\providecommand \BibitemShut  [1]{\csname bibitem#1\endcsname}%
\let\auto@bib@innerbib\@empty
\bibitem [{\citenamefont {Eremets}\ \emph {et~al.}(2008)\citenamefont
  {Eremets}, \citenamefont {Trojan}, \citenamefont {Medvedev}, \citenamefont
  {Tse},\ and\ \citenamefont {Yao}}]{Eremets_silane_2008}%
  \BibitemOpen
  \bibfield  {author} {\bibinfo {author} {\bibfnamefont {M.}~\bibnamefont
  {Eremets}}, \bibinfo {author} {\bibfnamefont {I.}~\bibnamefont {Trojan}},
  \bibinfo {author} {\bibfnamefont {S.}~\bibnamefont {Medvedev}}, \bibinfo
  {author} {\bibfnamefont {J.}~\bibnamefont {Tse}}, \ and\ \bibinfo {author}
  {\bibfnamefont {Y.}~\bibnamefont {Yao}},\ }\href@noop {} {\bibfield
  {journal} {\bibinfo  {journal} {Science}\ }\textbf {\bibinfo {volume}
  {319}},\ \bibinfo {pages} {1506} (\bibinfo {year} {2008})}\BibitemShut
  {NoStop}%
\bibitem [{\citenamefont {Drozdov}\ \emph {et~al.}(2015)\citenamefont
  {Drozdov}, \citenamefont {Eremets}, \citenamefont {Troyan}, \citenamefont
  {Ksenofontov},\ and\ \citenamefont {Shylin}}]{DrozdovEremets_Nature2015}%
  \BibitemOpen
  \bibfield  {author} {\bibinfo {author} {\bibfnamefont {A.~P.}\ \bibnamefont
  {Drozdov}}, \bibinfo {author} {\bibfnamefont {M.~I.}\ \bibnamefont
  {Eremets}}, \bibinfo {author} {\bibfnamefont {I.~A.}\ \bibnamefont {Troyan}},
  \bibinfo {author} {\bibfnamefont {V.}~\bibnamefont {Ksenofontov}}, \ and\
  \bibinfo {author} {\bibfnamefont {S.~I.}\ \bibnamefont {Shylin}},\ }\href
  {\doibase doi:10.1038/nature14964} {\bibfield  {journal} {\bibinfo  {journal}
  {Nature}\ }\textbf {\bibinfo {volume} {525}},\ \bibinfo {pages} {73}
  (\bibinfo {year} {2015})}\BibitemShut {NoStop}%
\bibitem [{\citenamefont {Einaga}\ \emph {et~al.}(2016)\citenamefont {Einaga},
  \citenamefont {Sakata}, \citenamefont {Ishikawa}, \citenamefont {Shimizu},
  \citenamefont {Eremets}, \citenamefont {Drozdov}, \citenamefont {Troyan},
  \citenamefont {Hirao},\ and\ \citenamefont
  {Ohishi}}]{Einaga_H3S-crystal_NatPhys-2016}%
  \BibitemOpen
  \bibfield  {author} {\bibinfo {author} {\bibfnamefont {M.}~\bibnamefont
  {Einaga}}, \bibinfo {author} {\bibfnamefont {M.}~\bibnamefont {Sakata}},
  \bibinfo {author} {\bibfnamefont {T.}~\bibnamefont {Ishikawa}}, \bibinfo
  {author} {\bibfnamefont {K.}~\bibnamefont {Shimizu}}, \bibinfo {author}
  {\bibfnamefont {M.~I.}\ \bibnamefont {Eremets}}, \bibinfo {author}
  {\bibfnamefont {A.~P.}\ \bibnamefont {Drozdov}}, \bibinfo {author}
  {\bibfnamefont {I.~A.}\ \bibnamefont {Troyan}}, \bibinfo {author}
  {\bibfnamefont {N.}~\bibnamefont {Hirao}}, \ and\ \bibinfo {author}
  {\bibfnamefont {Y.}~\bibnamefont {Ohishi}},\ }\href@noop {} {\bibfield
  {journal} {\bibinfo  {journal} {Nature Physics}\ } (\bibinfo {year}
  {2016})}\BibitemShut {NoStop}%
\bibitem [{\citenamefont {Somayazulu}\ \emph {et~al.}(2019)\citenamefont
  {Somayazulu}, \citenamefont {Ahart}, \citenamefont {Mishra}, \citenamefont
  {Geballe}, \citenamefont {Baldini}, \citenamefont {Meng}, \citenamefont
  {Struzhkin},\ and\ \citenamefont {Hemley}}]{Hemley-LaH10_PRL_2019}%
  \BibitemOpen
  \bibfield  {author} {\bibinfo {author} {\bibfnamefont {M.}~\bibnamefont
  {Somayazulu}}, \bibinfo {author} {\bibfnamefont {M.}~\bibnamefont {Ahart}},
  \bibinfo {author} {\bibfnamefont {A.~K.}\ \bibnamefont {Mishra}}, \bibinfo
  {author} {\bibfnamefont {Z.~M.}\ \bibnamefont {Geballe}}, \bibinfo {author}
  {\bibfnamefont {M.}~\bibnamefont {Baldini}}, \bibinfo {author} {\bibfnamefont
  {Y.}~\bibnamefont {Meng}}, \bibinfo {author} {\bibfnamefont {V.~V.}\
  \bibnamefont {Struzhkin}}, \ and\ \bibinfo {author} {\bibfnamefont {R.~J.}\
  \bibnamefont {Hemley}},\ }\href@noop {} {\bibfield  {journal} {\bibinfo
  {journal} {Phys. Rev. Lett.}\ }\textbf {\bibinfo {volume} {122}},\ \bibinfo
  {pages} {027001} (\bibinfo {year} {2019})}\BibitemShut {NoStop}%
\bibitem [{\citenamefont {Drozdov}\ \emph {et~al.}(2019)\citenamefont
  {Drozdov}, \citenamefont {Kong}, \citenamefont {Minkov}, \citenamefont
  {Besedin}, \citenamefont {Kuzovnikov}, \citenamefont {Mozaffari},
  \citenamefont {Balicas}, \citenamefont {Balakirev}, \citenamefont {Graf},
  \citenamefont {Prakapenka}, \citenamefont {Greenberg}, \citenamefont
  {Knyazev}, \citenamefont {Tkacz},\ and\ \citenamefont
  {Eremets}}]{Nature_LaH_Eremets_2019}%
  \BibitemOpen
  \bibfield  {author} {\bibinfo {author} {\bibfnamefont {A.~P.}\ \bibnamefont
  {Drozdov}}, \bibinfo {author} {\bibfnamefont {P.~P.}\ \bibnamefont {Kong}},
  \bibinfo {author} {\bibfnamefont {V.~S.}\ \bibnamefont {Minkov}}, \bibinfo
  {author} {\bibfnamefont {S.~P.}\ \bibnamefont {Besedin}}, \bibinfo {author}
  {\bibfnamefont {M.~A.}\ \bibnamefont {Kuzovnikov}}, \bibinfo {author}
  {\bibfnamefont {S.}~\bibnamefont {Mozaffari}}, \bibinfo {author}
  {\bibfnamefont {L.}~\bibnamefont {Balicas}}, \bibinfo {author} {\bibfnamefont
  {F.~F.}\ \bibnamefont {Balakirev}}, \bibinfo {author} {\bibfnamefont {D.~E.}\
  \bibnamefont {Graf}}, \bibinfo {author} {\bibfnamefont {V.~B.}\ \bibnamefont
  {Prakapenka}}, \bibinfo {author} {\bibfnamefont {E.}~\bibnamefont
  {Greenberg}}, \bibinfo {author} {\bibfnamefont {D.~A.}\ \bibnamefont
  {Knyazev}}, \bibinfo {author} {\bibfnamefont {M.}~\bibnamefont {Tkacz}}, \
  and\ \bibinfo {author} {\bibfnamefont {M.~I.}\ \bibnamefont {Eremets}},\
  }\href@noop {} {\bibfield  {journal} {\bibinfo  {journal} {Nature}\ }\textbf
  {\bibinfo {volume} {569}},\ \bibinfo {pages} {528} (\bibinfo {year}
  {2019})}\BibitemShut {NoStop}%
\bibitem [{\citenamefont {Errea}\ \emph {et~al.}(2020)\citenamefont {Errea},
  \citenamefont {Belli}, \citenamefont {Monacelli}, \citenamefont {Sanna},
  \citenamefont {Koretsune}, \citenamefont {Tadano}, \citenamefont {Bianco},
  \citenamefont {Calandra}, \citenamefont {Arita}, \citenamefont {Mauri},\ and\
  \citenamefont {Flores-Livas}}]{Errea_Nature_2020}%
  \BibitemOpen
  \bibfield  {author} {\bibinfo {author} {\bibfnamefont {I.}~\bibnamefont
  {Errea}}, \bibinfo {author} {\bibfnamefont {F.}~\bibnamefont {Belli}},
  \bibinfo {author} {\bibfnamefont {L.}~\bibnamefont {Monacelli}}, \bibinfo
  {author} {\bibfnamefont {A.}~\bibnamefont {Sanna}}, \bibinfo {author}
  {\bibfnamefont {T.}~\bibnamefont {Koretsune}}, \bibinfo {author}
  {\bibfnamefont {T.}~\bibnamefont {Tadano}}, \bibinfo {author} {\bibfnamefont
  {R.}~\bibnamefont {Bianco}}, \bibinfo {author} {\bibfnamefont
  {M.}~\bibnamefont {Calandra}}, \bibinfo {author} {\bibfnamefont
  {R.}~\bibnamefont {Arita}}, \bibinfo {author} {\bibfnamefont
  {F.}~\bibnamefont {Mauri}}, \ and\ \bibinfo {author} {\bibfnamefont {J.~A.}\
  \bibnamefont {Flores-Livas}},\ }\href@noop {} {\bibfield  {journal} {\bibinfo
   {journal} {Nature}\ }\textbf {\bibinfo {volume} {578}},\ \bibinfo {pages}
  {66} (\bibinfo {year} {2020})}\BibitemShut {NoStop}%
\bibitem [{\citenamefont {Flores-Livas}\ \emph
  {et~al.}(2020{\natexlab{a}})\citenamefont {Flores-Livas}, \citenamefont
  {Boeri}, \citenamefont {Sanna}, \citenamefont {Profeta}, \citenamefont
  {Arita},\ and\ \citenamefont {Eremets}}]{Flores_review}%
  \BibitemOpen
  \bibfield  {author} {\bibinfo {author} {\bibfnamefont {J.~A.}\ \bibnamefont
  {Flores-Livas}}, \bibinfo {author} {\bibfnamefont {L.}~\bibnamefont {Boeri}},
  \bibinfo {author} {\bibfnamefont {A.}~\bibnamefont {Sanna}}, \bibinfo
  {author} {\bibfnamefont {G.}~\bibnamefont {Profeta}}, \bibinfo {author}
  {\bibfnamefont {R.}~\bibnamefont {Arita}}, \ and\ \bibinfo {author}
  {\bibfnamefont {M.}~\bibnamefont {Eremets}},\ }\href {\doibase
  https://doi.org/10.1016/j.physrep.2020.02.003} {\bibfield  {journal}
  {\bibinfo  {journal} {Physics Reports}\ }\textbf {\bibinfo {volume} {856}},\
  \bibinfo {pages} {1} (\bibinfo {year} {2020}{\natexlab{a}})},\ \bibinfo
  {note} {a perspective on conventional high-temperature superconductors at
  high pressure: Methods and materials}\BibitemShut {NoStop}%
\bibitem [{\citenamefont {Bi}\ \emph {et~al.}(2019)\citenamefont {Bi},
  \citenamefont {Zarifi}, \citenamefont {Terpstra},\ and\ \citenamefont
  {Zurek}}]{BI2019}%
  \BibitemOpen
  \bibfield  {author} {\bibinfo {author} {\bibfnamefont {T.}~\bibnamefont
  {Bi}}, \bibinfo {author} {\bibfnamefont {N.}~\bibnamefont {Zarifi}}, \bibinfo
  {author} {\bibfnamefont {T.}~\bibnamefont {Terpstra}}, \ and\ \bibinfo
  {author} {\bibfnamefont {E.}~\bibnamefont {Zurek}},\ }in\ \href {\doibase
  https://doi.org/10.1016/B978-0-12-409547-2.11435-0} {\emph {\bibinfo
  {booktitle} {Reference Module in Chemistry, Molecular Sciences and Chemical
  Engineering}}}\ (\bibinfo  {publisher} {Elsevier},\ \bibinfo {year}
  {2019})\BibitemShut {NoStop}%
\bibitem [{\citenamefont {Pickard}\ \emph {et~al.}(2020)\citenamefont
  {Pickard}, \citenamefont {Errea},\ and\ \citenamefont
  {Eremets}}]{pickard2020superconducting}%
  \BibitemOpen
  \bibfield  {author} {\bibinfo {author} {\bibfnamefont {C.~J.}\ \bibnamefont
  {Pickard}}, \bibinfo {author} {\bibfnamefont {I.}~\bibnamefont {Errea}}, \
  and\ \bibinfo {author} {\bibfnamefont {M.~I.}\ \bibnamefont {Eremets}},\
  }\href@noop {} {\bibfield  {journal} {\bibinfo  {journal} {Annual Review of
  Condensed Matter Physics}\ }\textbf {\bibinfo {volume} {11}},\ \bibinfo
  {pages} {57} (\bibinfo {year} {2020})}\BibitemShut {NoStop}%
\bibitem [{\citenamefont {Zhang}\ \emph {et~al.}(2017)\citenamefont {Zhang},
  \citenamefont {Wang}, \citenamefont {Lv},\ and\ \citenamefont
  {Ma}}]{zhang2017materials}%
  \BibitemOpen
  \bibfield  {author} {\bibinfo {author} {\bibfnamefont {L.}~\bibnamefont
  {Zhang}}, \bibinfo {author} {\bibfnamefont {Y.}~\bibnamefont {Wang}},
  \bibinfo {author} {\bibfnamefont {J.}~\bibnamefont {Lv}}, \ and\ \bibinfo
  {author} {\bibfnamefont {Y.}~\bibnamefont {Ma}},\ }\href@noop {} {\bibfield
  {journal} {\bibinfo  {journal} {Nature Reviews Materials}\ }\textbf {\bibinfo
  {volume} {2}},\ \bibinfo {pages} {1} (\bibinfo {year} {2017})}\BibitemShut
  {NoStop}%
\bibitem [{\citenamefont {Oganov}\ \emph {et~al.}(2019)\citenamefont {Oganov},
  \citenamefont {Pickard}, \citenamefont {Zhu},\ and\ \citenamefont
  {Needs}}]{oganov2019structure}%
  \BibitemOpen
  \bibfield  {author} {\bibinfo {author} {\bibfnamefont {A.~R.}\ \bibnamefont
  {Oganov}}, \bibinfo {author} {\bibfnamefont {C.~J.}\ \bibnamefont {Pickard}},
  \bibinfo {author} {\bibfnamefont {Q.}~\bibnamefont {Zhu}}, \ and\ \bibinfo
  {author} {\bibfnamefont {R.~J.}\ \bibnamefont {Needs}},\ }\href@noop {}
  {\bibfield  {journal} {\bibinfo  {journal} {Nature Reviews Materials}\
  }\textbf {\bibinfo {volume} {4}},\ \bibinfo {pages} {331} (\bibinfo {year}
  {2019})}\BibitemShut {NoStop}%
\bibitem [{\citenamefont {Snider}\ \emph {et~al.}(2020)\citenamefont {Snider},
  \citenamefont {Dasenbrock-Gammon}, \citenamefont {McBride}, \citenamefont
  {Debessai}, \citenamefont {Vindana}, \citenamefont {Vencatasamy},
  \citenamefont {Lawler}, \citenamefont {Salamat},\ and\ \citenamefont
  {Dias}}]{Snider2020}%
  \BibitemOpen
  \bibfield  {author} {\bibinfo {author} {\bibfnamefont {E.}~\bibnamefont
  {Snider}}, \bibinfo {author} {\bibfnamefont {N.}~\bibnamefont
  {Dasenbrock-Gammon}}, \bibinfo {author} {\bibfnamefont {R.}~\bibnamefont
  {McBride}}, \bibinfo {author} {\bibfnamefont {M.}~\bibnamefont {Debessai}},
  \bibinfo {author} {\bibfnamefont {H.}~\bibnamefont {Vindana}}, \bibinfo
  {author} {\bibfnamefont {K.}~\bibnamefont {Vencatasamy}}, \bibinfo {author}
  {\bibfnamefont {K.~V.}\ \bibnamefont {Lawler}}, \bibinfo {author}
  {\bibfnamefont {A.}~\bibnamefont {Salamat}}, \ and\ \bibinfo {author}
  {\bibfnamefont {R.~P.}\ \bibnamefont {Dias}},\ }\href@noop {} {\bibfield
  {journal} {\bibinfo  {journal} {Nature}\ }\textbf {\bibinfo {volume} {586}},\
  \bibinfo {pages} {373} (\bibinfo {year} {2020})}\BibitemShut {NoStop}%
\bibitem [{\citenamefont {Hirsch}\ and\ \citenamefont
  {Marsiglio}(2020)}]{hirsch2020absence}%
  \BibitemOpen
  \bibfield  {author} {\bibinfo {author} {\bibfnamefont {J.~E.}\ \bibnamefont
  {Hirsch}}\ and\ \bibinfo {author} {\bibfnamefont {F.}~\bibnamefont
  {Marsiglio}},\ }\href@noop {} {\enquote {\bibinfo {title} {Absence of high
  temperature superconductivity in hydrides under pressure},}\ } (\bibinfo
  {year} {2020}),\ \Eprint {http://arxiv.org/abs/2010.10307} {arXiv:2010.10307
  [cond-mat.supr-con]} \BibitemShut {NoStop}%
\bibitem [{\citenamefont {Hirsch}\ and\ \citenamefont
  {Marsiglio}(2021{\natexlab{a}})}]{hirsch2021absence}%
  \BibitemOpen
  \bibfield  {author} {\bibinfo {author} {\bibfnamefont {J.}~\bibnamefont
  {Hirsch}}\ and\ \bibinfo {author} {\bibfnamefont {F.}~\bibnamefont
  {Marsiglio}},\ }\href@noop {} {\bibfield  {journal} {\bibinfo  {journal}
  {Physica C: Superconductivity and its Applications}\ ,\ \bibinfo {pages}
  {1353866}} (\bibinfo {year} {2021}{\natexlab{a}})}\BibitemShut {NoStop}%
\bibitem [{\citenamefont {Hirsch}\ and\ \citenamefont
  {Marsiglio}(2021{\natexlab{b}})}]{Hirsch_PRB_2021}%
  \BibitemOpen
  \bibfield  {author} {\bibinfo {author} {\bibfnamefont {J.~E.}\ \bibnamefont
  {Hirsch}}\ and\ \bibinfo {author} {\bibfnamefont {F.}~\bibnamefont
  {Marsiglio}},\ }\href {\doibase 10.1103/PhysRevB.103.134505} {\bibfield
  {journal} {\bibinfo  {journal} {Phys. Rev. B}\ }\textbf {\bibinfo {volume}
  {103}},\ \bibinfo {pages} {134505} (\bibinfo {year}
  {2021}{\natexlab{b}})}\BibitemShut {NoStop}%
\bibitem [{\citenamefont {Talantsev}(2021)}]{talantsev2021electron}%
  \BibitemOpen
  \bibfield  {author} {\bibinfo {author} {\bibfnamefont {E.~F.}\ \bibnamefont
  {Talantsev}},\ }\href@noop {} {\bibfield  {journal} {\bibinfo  {journal}
  {Superconductor Science and Technology}\ }\textbf {\bibinfo {volume} {34}},\
  \bibinfo {pages} {034001} (\bibinfo {year} {2021})}\BibitemShut {NoStop}%
\bibitem [{\citenamefont {Dogan}\ and\ \citenamefont
  {Cohen}(2021)}]{dogan2021anomalous}%
  \BibitemOpen
  \bibfield  {author} {\bibinfo {author} {\bibfnamefont {M.}~\bibnamefont
  {Dogan}}\ and\ \bibinfo {author} {\bibfnamefont {M.~L.}\ \bibnamefont
  {Cohen}},\ }\href@noop {} {\bibfield  {journal} {\bibinfo  {journal} {Physica
  C: Superconductivity and its Applications}\ ,\ \bibinfo {pages} {1353851}}
  (\bibinfo {year} {2021})}\BibitemShut {NoStop}%
\bibitem [{\citenamefont {Sun}\ \emph {et~al.}(2020)\citenamefont {Sun},
  \citenamefont {Tian}, \citenamefont {Jiang}, \citenamefont {Li},
  \citenamefont {Li}, \citenamefont {Iitaka}, \citenamefont {Zhong},\ and\
  \citenamefont {Xie}}]{Sun2020}%
  \BibitemOpen
  \bibfield  {author} {\bibinfo {author} {\bibfnamefont {Y.}~\bibnamefont
  {Sun}}, \bibinfo {author} {\bibfnamefont {Y.}~\bibnamefont {Tian}}, \bibinfo
  {author} {\bibfnamefont {B.}~\bibnamefont {Jiang}}, \bibinfo {author}
  {\bibfnamefont {X.}~\bibnamefont {Li}}, \bibinfo {author} {\bibfnamefont
  {H.}~\bibnamefont {Li}}, \bibinfo {author} {\bibfnamefont {T.}~\bibnamefont
  {Iitaka}}, \bibinfo {author} {\bibfnamefont {X.}~\bibnamefont {Zhong}}, \
  and\ \bibinfo {author} {\bibfnamefont {Y.}~\bibnamefont {Xie}},\ }\href
  {\doibase 10.1103/PhysRevB.101.174102} {\bibfield  {journal} {\bibinfo
  {journal} {Phys. Rev. B}\ }\textbf {\bibinfo {volume} {101}},\ \bibinfo
  {pages} {174102} (\bibinfo {year} {2020})}\BibitemShut {NoStop}%
\bibitem [{\citenamefont {Cui}\ \emph {et~al.}(2020)\citenamefont {Cui},
  \citenamefont {Bi}, \citenamefont {Shi}, \citenamefont {Li}, \citenamefont
  {Liu}, \citenamefont {Zurek},\ and\ \citenamefont {Hemley}}]{Cui2020}%
  \BibitemOpen
  \bibfield  {author} {\bibinfo {author} {\bibfnamefont {W.}~\bibnamefont
  {Cui}}, \bibinfo {author} {\bibfnamefont {T.}~\bibnamefont {Bi}}, \bibinfo
  {author} {\bibfnamefont {J.}~\bibnamefont {Shi}}, \bibinfo {author}
  {\bibfnamefont {Y.}~\bibnamefont {Li}}, \bibinfo {author} {\bibfnamefont
  {H.}~\bibnamefont {Liu}}, \bibinfo {author} {\bibfnamefont {E.}~\bibnamefont
  {Zurek}}, \ and\ \bibinfo {author} {\bibfnamefont {R.~J.}\ \bibnamefont
  {Hemley}},\ }\href {\doibase 10.1103/PhysRevB.101.134504} {\bibfield
  {journal} {\bibinfo  {journal} {Phys. Rev. B}\ }\textbf {\bibinfo {volume}
  {101}},\ \bibinfo {pages} {134504} (\bibinfo {year} {2020})}\BibitemShut
  {NoStop}%
\bibitem [{\citenamefont {Ge}\ \emph {et~al.}(2020)\citenamefont {Ge},
  \citenamefont {Zhang}, \citenamefont {Dias}, \citenamefont {Hemley},\ and\
  \citenamefont {Yao}}]{Ge_2020}%
  \BibitemOpen
  \bibfield  {author} {\bibinfo {author} {\bibfnamefont {Y.}~\bibnamefont
  {Ge}}, \bibinfo {author} {\bibfnamefont {F.}~\bibnamefont {Zhang}}, \bibinfo
  {author} {\bibfnamefont {R.~P.}\ \bibnamefont {Dias}}, \bibinfo {author}
  {\bibfnamefont {R.~J.}\ \bibnamefont {Hemley}}, \ and\ \bibinfo {author}
  {\bibfnamefont {Y.}~\bibnamefont {Yao}},\ }\href {\doibase
  10.1016/j.mtphys.2020.100330} {\bibfield  {journal} {\bibinfo  {journal}
  {Materials Today Physics}\ }\textbf {\bibinfo {volume} {15}},\ \bibinfo
  {pages} {100330} (\bibinfo {year} {2020})}\BibitemShut {NoStop}%
\bibitem [{\citenamefont {Hu}\ \emph {et~al.}(2020)\citenamefont {Hu},
  \citenamefont {Paul}, \citenamefont {Karasiev},\ and\ \citenamefont
  {Dias}}]{hu2020carbondoped}%
  \BibitemOpen
  \bibfield  {author} {\bibinfo {author} {\bibfnamefont {S.~X.}\ \bibnamefont
  {Hu}}, \bibinfo {author} {\bibfnamefont {R.}~\bibnamefont {Paul}}, \bibinfo
  {author} {\bibfnamefont {V.~V.}\ \bibnamefont {Karasiev}}, \ and\ \bibinfo
  {author} {\bibfnamefont {R.~P.}\ \bibnamefont {Dias}},\ }\href@noop {}
  {\enquote {\bibinfo {title} {{Carbon-Doped Sulfur Hydrides as
  Room-Temperature Superconductors at 270 GPa}},}\ } (\bibinfo {year} {2020}),\
  \Eprint {http://arxiv.org/abs/2012.10259} {arXiv:2012.10259
  [cond-mat.supr-con]} \BibitemShut {NoStop}%
\bibitem [{Note1()}]{Note1}%
  \BibitemOpen
  \bibinfo {note} {The compositional and configurational space of C-S-H was
  explored using the minima-hopping method~\cite
  {Goedecker_mhm_2004,Amsler_mhm_2010,flores-sanna_PH3_2016,flores2020crystal}.
  The representative compositions of the ternary phase diagram were carried out
  with different formula units (containing up to 48 atoms) and carried out
  under pressure. Energy, atomic forces, and stresses were evaluated using the
  density-functional theory (further details are included in the Supplemental
  Information).}\BibitemShut {Stop}%
\bibitem [{\citenamefont {Flores-Livas}\ \emph
  {et~al.}(2020{\natexlab{b}})\citenamefont {Flores-Livas}, \citenamefont
  {Wang}, \citenamefont {Nomoto}, \citenamefont {Koretsune}, \citenamefont
  {Ma}, \citenamefont {Arita},\ and\ \citenamefont
  {Eremets}}]{LaAlH10_Rome-Tokyo}%
  \BibitemOpen
  \bibfield  {author} {\bibinfo {author} {\bibfnamefont {J.~A.}\ \bibnamefont
  {Flores-Livas}}, \bibinfo {author} {\bibfnamefont {T.}~\bibnamefont {Wang}},
  \bibinfo {author} {\bibfnamefont {T.}~\bibnamefont {Nomoto}}, \bibinfo
  {author} {\bibfnamefont {T.}~\bibnamefont {Koretsune}}, \bibinfo {author}
  {\bibfnamefont {Y.}~\bibnamefont {Ma}}, \bibinfo {author} {\bibfnamefont
  {R.}~\bibnamefont {Arita}}, \ and\ \bibinfo {author} {\bibfnamefont
  {M.}~\bibnamefont {Eremets}},\ }\href@noop {} {\bibfield  {journal} {\bibinfo
   {journal} {arXiv preprint arXiv:2010.06446}\ } (\bibinfo {year}
  {2020}{\natexlab{b}})}\BibitemShut {NoStop}%
\bibitem [{\citenamefont {Akashi}(2020)}]{PRB_VHA_Akashi}%
  \BibitemOpen
  \bibfield  {author} {\bibinfo {author} {\bibfnamefont {R.}~\bibnamefont
  {Akashi}},\ }\href {\doibase 10.1103/PhysRevB.101.075126} {\bibfield
  {journal} {\bibinfo  {journal} {Phys. Rev. B}\ }\textbf {\bibinfo {volume}
  {101}},\ \bibinfo {pages} {075126} (\bibinfo {year} {2020})}\BibitemShut
  {NoStop}%
\bibitem [{\citenamefont {McMillan}(1968)}]{McMillanTC_PR_1968}%
  \BibitemOpen
  \bibfield  {author} {\bibinfo {author} {\bibfnamefont {W.~L.}\ \bibnamefont
  {McMillan}},\ }\href {\doibase 10.1103/PhysRev.167.331} {\bibfield  {journal}
  {\bibinfo  {journal} {Phys. Rev.}\ }\textbf {\bibinfo {volume} {167}},\
  \bibinfo {pages} {331} (\bibinfo {year} {1968})}\BibitemShut {NoStop}%
\bibitem [{\citenamefont {Allen}\ and\ \citenamefont
  {Dynes}(1975)}]{AllenDynes_PRB1975}%
  \BibitemOpen
  \bibfield  {author} {\bibinfo {author} {\bibfnamefont {P.~B.}\ \bibnamefont
  {Allen}}\ and\ \bibinfo {author} {\bibfnamefont {R.~C.}\ \bibnamefont
  {Dynes}},\ }\href {\doibase 10.1103/PhysRevB.12.905} {\bibfield  {journal}
  {\bibinfo  {journal} {Phys. Rev. B}\ }\textbf {\bibinfo {volume} {12}},\
  \bibinfo {pages} {905} (\bibinfo {year} {1975})}\BibitemShut {NoStop}%
\bibitem [{\citenamefont {Allen}\ and\ \citenamefont {Mitrovi{\'
  c}}(1983)}]{AllenMitrovic1983}%
  \BibitemOpen
  \bibfield  {author} {\bibinfo {author} {\bibfnamefont {P.~B.}\ \bibnamefont
  {Allen}}\ and\ \bibinfo {author} {\bibfnamefont {B.}~\bibnamefont {Mitrovi{\'
  c}}},\ }\href {\doibase http://dx.doi.org/10.1016/S0081-1947(08)60665-7}
  {\emph {\bibinfo {title} {Theory of Superconducting Tc}}},\ \bibinfo {series}
  {Solid State Physics}, Vol.~\bibinfo {volume} {37}\ (\bibinfo  {publisher}
  {Academic Press},\ \bibinfo {year} {1983})\ pp.\ \bibinfo {pages} {1 --
  92}\BibitemShut {NoStop}%
\bibitem [{\citenamefont {Bl\"ochl}\ \emph {et~al.}(1994)\citenamefont
  {Bl\"ochl}, \citenamefont {Jepsen},\ and\ \citenamefont
  {Andersen}}]{Blochl_1994}%
  \BibitemOpen
  \bibfield  {author} {\bibinfo {author} {\bibfnamefont {P.~E.}\ \bibnamefont
  {Bl\"ochl}}, \bibinfo {author} {\bibfnamefont {O.}~\bibnamefont {Jepsen}}, \
  and\ \bibinfo {author} {\bibfnamefont {O.~K.}\ \bibnamefont {Andersen}},\
  }\href {\doibase 10.1103/PhysRevB.49.16223} {\bibfield  {journal} {\bibinfo
  {journal} {Phys. Rev. B}\ }\textbf {\bibinfo {volume} {49}},\ \bibinfo
  {pages} {16223} (\bibinfo {year} {1994})}\BibitemShut {NoStop}%
\bibitem [{\citenamefont {Koretsune}\ and\ \citenamefont
  {Arita}(2017)}]{Koretsune_2017}%
  \BibitemOpen
  \bibfield  {author} {\bibinfo {author} {\bibfnamefont {T.}~\bibnamefont
  {Koretsune}}\ and\ \bibinfo {author} {\bibfnamefont {R.}~\bibnamefont
  {Arita}},\ }\href@noop {} {\bibfield  {journal} {\bibinfo  {journal}
  {Computer Physics Communications}\ }\textbf {\bibinfo {volume} {220}},\
  \bibinfo {pages} {239} (\bibinfo {year} {2017})}\BibitemShut {NoStop}%
\bibitem [{\citenamefont {Wang}\ \emph {et~al.}(2020)\citenamefont {Wang},
  \citenamefont {Nomoto}, \citenamefont {Nomura}, \citenamefont {Shinaoka},
  \citenamefont {Otsuki}, \citenamefont {Koretsune},\ and\ \citenamefont
  {Arita}}]{Wang2020}%
  \BibitemOpen
  \bibfield  {author} {\bibinfo {author} {\bibfnamefont {T.}~\bibnamefont
  {Wang}}, \bibinfo {author} {\bibfnamefont {T.}~\bibnamefont {Nomoto}},
  \bibinfo {author} {\bibfnamefont {Y.}~\bibnamefont {Nomura}}, \bibinfo
  {author} {\bibfnamefont {H.}~\bibnamefont {Shinaoka}}, \bibinfo {author}
  {\bibfnamefont {J.}~\bibnamefont {Otsuki}}, \bibinfo {author} {\bibfnamefont
  {T.}~\bibnamefont {Koretsune}}, \ and\ \bibinfo {author} {\bibfnamefont
  {R.}~\bibnamefont {Arita}},\ }\href {\doibase 10.1103/PhysRevB.102.134503}
  {\bibfield  {journal} {\bibinfo  {journal} {Phys. Rev. B}\ }\textbf {\bibinfo
  {volume} {102}},\ \bibinfo {pages} {134503} (\bibinfo {year}
  {2020})}\BibitemShut {NoStop}%
\bibitem [{\citenamefont {Sano}\ \emph {et~al.}(2016)\citenamefont {Sano},
  \citenamefont {Koretsune}, \citenamefont {Tadano}, \citenamefont {Akashi},\
  and\ \citenamefont {Arita}}]{Sano2016}%
  \BibitemOpen
  \bibfield  {author} {\bibinfo {author} {\bibfnamefont {W.}~\bibnamefont
  {Sano}}, \bibinfo {author} {\bibfnamefont {T.}~\bibnamefont {Koretsune}},
  \bibinfo {author} {\bibfnamefont {T.}~\bibnamefont {Tadano}}, \bibinfo
  {author} {\bibfnamefont {R.}~\bibnamefont {Akashi}}, \ and\ \bibinfo {author}
  {\bibfnamefont {R.}~\bibnamefont {Arita}},\ }\href {\doibase
  10.1103/PhysRevB.93.094525} {\bibfield  {journal} {\bibinfo  {journal} {Phys.
  Rev. B}\ }\textbf {\bibinfo {volume} {93}},\ \bibinfo {pages} {094525}
  (\bibinfo {year} {2016})}\BibitemShut {NoStop}%
\bibitem [{Note2()}]{Note2}%
  \BibitemOpen
  \bibinfo {note} {To study the doping effect of C atoms in
  C$_x$S$_{1-x}$H$_3$, we used I) the virtual crystal approximation (VCA)~\cite
  {Nordheim_1931,Bellaiche&Vanderbilt_2000} and performed a thorough
  examination of C$_x$S$_{1-x}$H$_3$ with $x$ ranging from $1\%$ to $8.3\%$, in
  which for the virtual atom we use the potential $V_{\protect \rm VCA} = x
  V_{\protect \rm C}+(1-x)V_{\protect \rm S}$; II) the actual supercell
  calculation for C atoms doping with 8\% ratio. We use {\protect \textsc
  {Quantum ESPRESSO}}\protect \xspace ~\cite {Quantumespresso_2017} for the DFT
  and DFPT calculations~\cite {Baroni_DFPT_RMP2001}, with the
  exchange-correlation functional proposed by Perdew, Burke, and
  Ernzerhof~\cite {PBE_PRL1996}. For C and S atoms, the optimized
  norm-conserving Vanderbilt pseudopotential (ONCVPSP)~\cite {Hamann_2013} was
  employed. For H atom, we use the ultrasoft pseudopotentials~\cite
  {Vanderbilt_1990} provided in PSLibrary~\cite {PSLibrary_1.0.0}. To generate
  $V_{\protect \rm VCA}$, we use the tool \protect \textit {virtualv2.x}
  implemented in {\protect \textsc {Quantum ESPRESSO}}\protect \xspace . For
  the DFT calculation, the cutoff energy of the plane waves for wavefunction
  expansion is set to be 80 Ry, and cutoff of the charge density is 320 Ry. For
  calculations using $V_{\protect \rm VCA}$, we use an 18$\times $18$\times $18
  $\protect \bm {k}$-mesh and a 9$\times $9$\times $9 $\protect \bm {q}$-mesh
  for the phonon calculation and a 36$\times $36$\times $36 $\protect \bm
  {k}$-mesh for the electron-phonon calculation. For the electron-phonon
  calculation and the Eliashberg calculation, the electronic eigenenergies and
  wave-functions are calculated using a 36$\times $36$\times $36 $\protect \bm
  {k}$-mesh. We have also performed supercell calculation for
  C$_x$S$_{1-x}$H$_3$ with $x=0.083$, using a 3$\times $2$\times $2 supercell.
  The supercell calculation is performed on a 6$\times $6$\times $6 $\protect
  \bm {k}$-mesh and a 3$\times $3$\times $3 $\protect \bm {q}$-mesh for the
  DFPT calculation, and a 12$\times $12$\times $12 $\protect \bm {k}$-mesh is
  employed for the electron-phonon interaction and Eliashberg calculation. In
  the Eliashberg calculation, we use an RPA-type static Coulomb kernel, with a
  6$\times $6$\times $6 $\protect \bm {k}$-mesh and a 3$\times $3$\times $3
  $\protect \bm {q}$-mesh for the Coulomb calculation.}\BibitemShut {Stop}%
\bibitem [{\citenamefont {Flores-Livas}\ \emph
  {et~al.}(2016{\natexlab{a}})\citenamefont {Flores-Livas}, \citenamefont
  {Sanna},\ and\ \citenamefont {Gross}}]{flores-sanna_HSe_2016}%
  \BibitemOpen
  \bibfield  {author} {\bibinfo {author} {\bibfnamefont {A.~J.}\ \bibnamefont
  {Flores-Livas}}, \bibinfo {author} {\bibfnamefont {A.}~\bibnamefont {Sanna}},
  \ and\ \bibinfo {author} {\bibfnamefont {E.}~\bibnamefont {Gross}},\ }\href
  {\doibase 10.1140/epjb/e2016-70020-0} {\bibfield  {journal} {\bibinfo
  {journal} {Eur. Phys. J. B}\ }\textbf {\bibinfo {volume} {89}},\ \bibinfo
  {pages} {1} (\bibinfo {year} {2016}{\natexlab{a}})}\BibitemShut {NoStop}%
\bibitem [{\citenamefont {Akashi}\ \emph {et~al.}(2015)\citenamefont {Akashi},
  \citenamefont {Kawamura}, \citenamefont {Tsuneyuki}, \citenamefont {Nomura},\
  and\ \citenamefont {Arita}}]{Akashi2015}%
  \BibitemOpen
  \bibfield  {author} {\bibinfo {author} {\bibfnamefont {R.}~\bibnamefont
  {Akashi}}, \bibinfo {author} {\bibfnamefont {M.}~\bibnamefont {Kawamura}},
  \bibinfo {author} {\bibfnamefont {S.}~\bibnamefont {Tsuneyuki}}, \bibinfo
  {author} {\bibfnamefont {Y.}~\bibnamefont {Nomura}}, \ and\ \bibinfo {author}
  {\bibfnamefont {R.}~\bibnamefont {Arita}},\ }\href {\doibase
  10.1103/PhysRevB.91.224513} {\bibfield  {journal} {\bibinfo  {journal} {Phys.
  Rev. B}\ }\textbf {\bibinfo {volume} {91}},\ \bibinfo {pages} {224513}
  (\bibinfo {year} {2015})}\BibitemShut {NoStop}%
\bibitem [{\citenamefont {Akashi}\ \emph {et~al.}(2016)\citenamefont {Akashi},
  \citenamefont {Sano}, \citenamefont {Arita},\ and\ \citenamefont
  {Tsuneyuki}}]{Akashi2016}%
  \BibitemOpen
  \bibfield  {author} {\bibinfo {author} {\bibfnamefont {R.}~\bibnamefont
  {Akashi}}, \bibinfo {author} {\bibfnamefont {W.}~\bibnamefont {Sano}},
  \bibinfo {author} {\bibfnamefont {R.}~\bibnamefont {Arita}}, \ and\ \bibinfo
  {author} {\bibfnamefont {S.}~\bibnamefont {Tsuneyuki}},\ }\href {\doibase
  10.1103/PhysRevLett.117.075503} {\bibfield  {journal} {\bibinfo  {journal}
  {Phys. Rev. Lett.}\ }\textbf {\bibinfo {volume} {117}},\ \bibinfo {pages}
  {075503} (\bibinfo {year} {2016})}\BibitemShut {NoStop}%
\bibitem [{\citenamefont {Errea}\ \emph {et~al.}(2016)\citenamefont {Errea},
  \citenamefont {Calandra}, \citenamefont {Pickard}, \citenamefont {Nelson},
  \citenamefont {Needs}, \citenamefont {Li}, \citenamefont {Liu}, \citenamefont
  {Zhang}, \citenamefont {Ma},\ and\ \citenamefont
  {Mauri}}]{Nature_Errea_2016}%
  \BibitemOpen
  \bibfield  {author} {\bibinfo {author} {\bibfnamefont {I.}~\bibnamefont
  {Errea}}, \bibinfo {author} {\bibfnamefont {M.}~\bibnamefont {Calandra}},
  \bibinfo {author} {\bibfnamefont {C.~J.}\ \bibnamefont {Pickard}}, \bibinfo
  {author} {\bibfnamefont {J.~R.}\ \bibnamefont {Nelson}}, \bibinfo {author}
  {\bibfnamefont {R.~J.}\ \bibnamefont {Needs}}, \bibinfo {author}
  {\bibfnamefont {Y.}~\bibnamefont {Li}}, \bibinfo {author} {\bibfnamefont
  {H.}~\bibnamefont {Liu}}, \bibinfo {author} {\bibfnamefont {Y.}~\bibnamefont
  {Zhang}}, \bibinfo {author} {\bibfnamefont {Y.}~\bibnamefont {Ma}}, \ and\
  \bibinfo {author} {\bibfnamefont {F.}~\bibnamefont {Mauri}},\ }\href@noop {}
  {\bibfield  {journal} {\bibinfo  {journal} {Nature}\ }\textbf {\bibinfo
  {volume} {532}},\ \bibinfo {pages} {81} (\bibinfo {year} {2016})}\BibitemShut
  {NoStop}%
\bibitem [{\citenamefont {Minkov}\ \emph {et~al.}(2020)\citenamefont {Minkov},
  \citenamefont {Prakapenka}, \citenamefont {Greenberg},\ and\ \citenamefont
  {Eremets}}]{minkov2020boosted}%
  \BibitemOpen
  \bibfield  {author} {\bibinfo {author} {\bibfnamefont {V.~S.}\ \bibnamefont
  {Minkov}}, \bibinfo {author} {\bibfnamefont {V.~B.}\ \bibnamefont
  {Prakapenka}}, \bibinfo {author} {\bibfnamefont {E.}~\bibnamefont
  {Greenberg}}, \ and\ \bibinfo {author} {\bibfnamefont {M.~I.}\ \bibnamefont
  {Eremets}},\ }\href@noop {} {\bibfield  {journal} {\bibinfo  {journal}
  {Angewandte Chemie International Edition}\ }\textbf {\bibinfo {volume}
  {59}},\ \bibinfo {pages} {18970} (\bibinfo {year} {2020})}\BibitemShut
  {NoStop}%
\bibitem [{\citenamefont {Mozaffari}\ \emph {et~al.}(2019)\citenamefont
  {Mozaffari}, \citenamefont {Sun}, \citenamefont {Minkov}, \citenamefont
  {Drozdov}, \citenamefont {Knyazev}, \citenamefont {Betts}, \citenamefont
  {Einaga}, \citenamefont {Shimizu}, \citenamefont {Eremets}, \citenamefont
  {Balicas} \emph {et~al.}}]{mozaffari2019superconducting}%
  \BibitemOpen
  \bibfield  {author} {\bibinfo {author} {\bibfnamefont {S.}~\bibnamefont
  {Mozaffari}}, \bibinfo {author} {\bibfnamefont {D.}~\bibnamefont {Sun}},
  \bibinfo {author} {\bibfnamefont {V.~S.}\ \bibnamefont {Minkov}}, \bibinfo
  {author} {\bibfnamefont {A.~P.}\ \bibnamefont {Drozdov}}, \bibinfo {author}
  {\bibfnamefont {D.}~\bibnamefont {Knyazev}}, \bibinfo {author} {\bibfnamefont
  {J.~B.}\ \bibnamefont {Betts}}, \bibinfo {author} {\bibfnamefont
  {M.}~\bibnamefont {Einaga}}, \bibinfo {author} {\bibfnamefont
  {K.}~\bibnamefont {Shimizu}}, \bibinfo {author} {\bibfnamefont {M.~I.}\
  \bibnamefont {Eremets}}, \bibinfo {author} {\bibfnamefont {L.}~\bibnamefont
  {Balicas}},  \emph {et~al.},\ }\href@noop {} {\bibfield  {journal} {\bibinfo
  {journal} {Nature communications}\ }\textbf {\bibinfo {volume} {10}},\
  \bibinfo {pages} {1} (\bibinfo {year} {2019})}\BibitemShut {NoStop}%
\bibitem [{\citenamefont {Goedecker}(2004)}]{Goedecker_mhm_2004}%
  \BibitemOpen
  \bibfield  {author} {\bibinfo {author} {\bibfnamefont {S.}~\bibnamefont
  {Goedecker}},\ }\href@noop {} {\bibfield  {journal} {\bibinfo  {journal} {The
  Journal of Chemical Physics}\ }\textbf {\bibinfo {volume} {120}},\ \bibinfo
  {pages} {9911} (\bibinfo {year} {2004})}\BibitemShut {NoStop}%
\bibitem [{\citenamefont {Amsler}\ and\ \citenamefont
  {Goedecker}(2010)}]{Amsler_mhm_2010}%
  \BibitemOpen
  \bibfield  {author} {\bibinfo {author} {\bibfnamefont {M.}~\bibnamefont
  {Amsler}}\ and\ \bibinfo {author} {\bibfnamefont {S.}~\bibnamefont
  {Goedecker}},\ }\href@noop {} {\bibfield  {journal} {\bibinfo  {journal} {The
  Journal of Chemical Physics}\ }\textbf {\bibinfo {volume} {133}},\ \bibinfo
  {pages} {224104} (\bibinfo {year} {2010})}\BibitemShut {NoStop}%
\bibitem [{\citenamefont {Flores-Livas}\ \emph
  {et~al.}(2016{\natexlab{b}})\citenamefont {Flores-Livas}, \citenamefont
  {Amsler}, \citenamefont {Heil}, \citenamefont {Sanna}, \citenamefont {Boeri},
  \citenamefont {Profeta}, \citenamefont {Wolverton}, \citenamefont
  {Goedecker},\ and\ \citenamefont {Gross}}]{flores-sanna_PH3_2016}%
  \BibitemOpen
  \bibfield  {author} {\bibinfo {author} {\bibfnamefont {J.~A.}\ \bibnamefont
  {Flores-Livas}}, \bibinfo {author} {\bibfnamefont {M.}~\bibnamefont
  {Amsler}}, \bibinfo {author} {\bibfnamefont {C.}~\bibnamefont {Heil}},
  \bibinfo {author} {\bibfnamefont {A.}~\bibnamefont {Sanna}}, \bibinfo
  {author} {\bibfnamefont {L.}~\bibnamefont {Boeri}}, \bibinfo {author}
  {\bibfnamefont {G.}~\bibnamefont {Profeta}}, \bibinfo {author} {\bibfnamefont
  {C.}~\bibnamefont {Wolverton}}, \bibinfo {author} {\bibfnamefont
  {S.}~\bibnamefont {Goedecker}}, \ and\ \bibinfo {author} {\bibfnamefont
  {E.~K.~U.}\ \bibnamefont {Gross}},\ }\href {\doibase
  10.1103/PhysRevB.93.020508} {\bibfield  {journal} {\bibinfo  {journal} {Phys.
  Rev. B}\ }\textbf {\bibinfo {volume} {93}},\ \bibinfo {pages} {020508(R)}
  (\bibinfo {year} {2016}{\natexlab{b}})}\BibitemShut {NoStop}%
\bibitem [{\citenamefont {Flores-Livas}(2020)}]{flores2020crystal}%
  \BibitemOpen
  \bibfield  {author} {\bibinfo {author} {\bibfnamefont {J.~A.}\ \bibnamefont
  {Flores-Livas}},\ }\href@noop {} {\bibfield  {journal} {\bibinfo  {journal}
  {Journal of Physics: Condensed Matter}\ }\textbf {\bibinfo {volume} {32}},\
  \bibinfo {pages} {294002} (\bibinfo {year} {2020})}\BibitemShut {NoStop}%
\bibitem [{\citenamefont {Nordheim}(1931)}]{Nordheim_1931}%
  \BibitemOpen
  \bibfield  {author} {\bibinfo {author} {\bibfnamefont {L.}~\bibnamefont
  {Nordheim}},\ }\href@noop {} {\bibfield  {journal} {\bibinfo  {journal}
  {Annalen der Physik}\ }\textbf {\bibinfo {volume} {401}},\ \bibinfo {pages}
  {607} (\bibinfo {year} {1931})}\BibitemShut {NoStop}%
\bibitem [{\citenamefont {Bellaiche}\ and\ \citenamefont
  {Vanderbilt}(2000)}]{Bellaiche&Vanderbilt_2000}%
  \BibitemOpen
  \bibfield  {author} {\bibinfo {author} {\bibfnamefont {L.}~\bibnamefont
  {Bellaiche}}\ and\ \bibinfo {author} {\bibfnamefont {D.}~\bibnamefont
  {Vanderbilt}},\ }\href {\doibase 10.1103/PhysRevB.61.7877} {\bibfield
  {journal} {\bibinfo  {journal} {Phys. Rev. B}\ }\textbf {\bibinfo {volume}
  {61}},\ \bibinfo {pages} {7877} (\bibinfo {year} {2000})}\BibitemShut
  {NoStop}%
\bibitem [{\citenamefont {Giannozzi}\ \emph {et~al.}(2017)\citenamefont
  {Giannozzi}, \citenamefont {Andreussi}, \citenamefont {Brumme}, \citenamefont
  {Bunau}, \citenamefont {Nardelli}, \citenamefont {Calandra}, \citenamefont
  {Car}, \citenamefont {Cavazzoni}, \citenamefont {Ceresoli}, \citenamefont
  {Cococcioni} \emph {et~al.}}]{Quantumespresso_2017}%
  \BibitemOpen
  \bibfield  {author} {\bibinfo {author} {\bibfnamefont {P.}~\bibnamefont
  {Giannozzi}}, \bibinfo {author} {\bibfnamefont {O.}~\bibnamefont
  {Andreussi}}, \bibinfo {author} {\bibfnamefont {T.}~\bibnamefont {Brumme}},
  \bibinfo {author} {\bibfnamefont {O.}~\bibnamefont {Bunau}}, \bibinfo
  {author} {\bibfnamefont {M.~B.}\ \bibnamefont {Nardelli}}, \bibinfo {author}
  {\bibfnamefont {M.}~\bibnamefont {Calandra}}, \bibinfo {author}
  {\bibfnamefont {R.}~\bibnamefont {Car}}, \bibinfo {author} {\bibfnamefont
  {C.}~\bibnamefont {Cavazzoni}}, \bibinfo {author} {\bibfnamefont
  {D.}~\bibnamefont {Ceresoli}}, \bibinfo {author} {\bibfnamefont
  {M.}~\bibnamefont {Cococcioni}},  \emph {et~al.},\ }\href@noop {} {\bibfield
  {journal} {\bibinfo  {journal} {Journal of Physics: Condensed Matter}\
  }\textbf {\bibinfo {volume} {29}},\ \bibinfo {pages} {465901} (\bibinfo
  {year} {2017})}\BibitemShut {NoStop}%
\bibitem [{\citenamefont {Baroni}\ \emph {et~al.}(2001)\citenamefont {Baroni},
  \citenamefont {de~Gironcoli}, \citenamefont {Dal~Corso},\ and\ \citenamefont
  {Giannozzi}}]{Baroni_DFPT_RMP2001}%
  \BibitemOpen
  \bibfield  {author} {\bibinfo {author} {\bibfnamefont {S.}~\bibnamefont
  {Baroni}}, \bibinfo {author} {\bibfnamefont {S.}~\bibnamefont
  {de~Gironcoli}}, \bibinfo {author} {\bibfnamefont {A.}~\bibnamefont
  {Dal~Corso}}, \ and\ \bibinfo {author} {\bibfnamefont {P.}~\bibnamefont
  {Giannozzi}},\ }\href {\doibase 10.1103/RevModPhys.73.515} {\bibfield
  {journal} {\bibinfo  {journal} {Rev. Mod. Phys.}\ }\textbf {\bibinfo {volume}
  {73}},\ \bibinfo {pages} {515} (\bibinfo {year} {2001})}\BibitemShut
  {NoStop}%
\bibitem [{\citenamefont {Perdew}\ \emph {et~al.}(1996)\citenamefont {Perdew},
  \citenamefont {Burke},\ and\ \citenamefont {Ernzerhof}}]{PBE_PRL1996}%
  \BibitemOpen
  \bibfield  {author} {\bibinfo {author} {\bibfnamefont {J.~P.}\ \bibnamefont
  {Perdew}}, \bibinfo {author} {\bibfnamefont {K.}~\bibnamefont {Burke}}, \
  and\ \bibinfo {author} {\bibfnamefont {M.}~\bibnamefont {Ernzerhof}},\ }\href
  {\doibase 10.1103/PhysRevLett.77.3865} {\bibfield  {journal} {\bibinfo
  {journal} {Phys. Rev. Lett.}\ }\textbf {\bibinfo {volume} {77}},\ \bibinfo
  {pages} {3865} (\bibinfo {year} {1996})}\BibitemShut {NoStop}%
\bibitem [{\citenamefont {Hamann}(2013)}]{Hamann_2013}%
  \BibitemOpen
  \bibfield  {author} {\bibinfo {author} {\bibfnamefont {D.~R.}\ \bibnamefont
  {Hamann}},\ }\href {\doibase 10.1103/PhysRevB.88.085117} {\bibfield
  {journal} {\bibinfo  {journal} {Phys. Rev. B}\ }\textbf {\bibinfo {volume}
  {88}},\ \bibinfo {pages} {085117} (\bibinfo {year} {2013})}\BibitemShut
  {NoStop}%
\bibitem [{\citenamefont {Vanderbilt}(1990)}]{Vanderbilt_1990}%
  \BibitemOpen
  \bibfield  {author} {\bibinfo {author} {\bibfnamefont {D.}~\bibnamefont
  {Vanderbilt}},\ }\href {\doibase 10.1103/PhysRevB.41.7892} {\bibfield
  {journal} {\bibinfo  {journal} {Phys. Rev. B}\ }\textbf {\bibinfo {volume}
  {41}},\ \bibinfo {pages} {7892} (\bibinfo {year} {1990})}\BibitemShut
  {NoStop}%
\bibitem [{\citenamefont {{Dal Corso}}(2014)}]{PSLibrary_1.0.0}%
  \BibitemOpen
  \bibfield  {author} {\bibinfo {author} {\bibfnamefont {A.}~\bibnamefont {{Dal
  Corso}}},\ }\href {\doibase https://doi.org/10.1016/j.commatsci.2014.07.043}
  {\bibfield  {journal} {\bibinfo  {journal} {Computational Materials Science}\
  }\textbf {\bibinfo {volume} {95}},\ \bibinfo {pages} {337 } (\bibinfo {year}
  {2014})}\BibitemShut {NoStop}%
\end{thebibliography}%
 
\end{document}